\documentclass[12,preprint]{aastex}
\usepackage{graphicx}
\usepackage{natbib}

\shorttitle{OUTFLOWS AND FEEDBACK IN MRK573} 
\shortauthors{SCHLESINGER ET AL.}

\slugcomment{ApJ in preparation [22 January 2009]}

\newcommand{\hst}{{\it HST\,\,}}

\newcommand{\kms}{km s$^{-1}$}

\newcommand{\oiii}{[\ion{O}{3}]}

\begin{document}

\title{The Nuclear Outflows and Feedback in the Seyfert 2 Galaxy 
Markarian 573\altaffilmark{1}}

\altaffiltext{1}{Based on observations with the
NASA/ESA {\it Hubble Space Telescope} obtained at the the Space Telescope
Science Institute, which is operated by the Association of Universities for
Research in Astronomy, Incorporated, under NASA contract NAS5-26555.}

\author{K. Schlesinger, Richard W. Pogge\altaffilmark{2}, Paul Martini\altaffilmark{2}}

\altaffiltext{2}{Center for Cosmology and AstroParticle Physics; The Ohio State University; 191 West Woodruff Avenue; Columbus, OH 43210} 

\affil{Department of Astronomy; The Ohio State University; 140 W. 18th Avenue, 
Columbus, OH 43210; kschles@astronomy.ohio-state.edu} 

\author{Joseph C. Shields}
\affil{Physics \& Astronomy Department; Ohio University; Athens, OH 45701}

\author{Dale Fields}

\affil{Department of Physics and Planetary Sciences; Pierce College; Woodland Hills, California 91371} 

\begin{abstract}
We present a study of outflow and feedback in the well-known Seyfert~2
galaxy Markarian 573 using high angular resolution long-slit
spectrophotometry obtained with the \emph{Hubble Space Telescope}
Imaging Spectrograph (STIS).  Through analysis of the kinematics and ionization
state of a biconical outflow region emanating from the nucleus, we find
that the outflow does not significantly accelerate the surrounding
host-galaxy interstellar gas and is too weak to be a strong ionization 
mechanism in the extended emission regions.  Instead, the excitation of the extended
regions is consistent with photoionization by the active nucleus.  From
energetics arguments we show that the nuclear outflow is slow and heavy
and has a mechanical luminosity that is only $\sim$1\% of the estimated
bolometric luminosity of the system.  The energy in the outflow is able
to mildly shape the gas in the extended regions but appears to be
insufficient to unbind it, or even to 
plausibly disrupt star formation.  These results are at odds with the
picture of strong AGN feedback that has been invoked to explain certain
aspects of galaxy evolution.

\end{abstract}


\section{Introduction}

Recent studies of galaxy evolution have invoked a significant component
of energetic feedback from an Active Galactic Nucleus (AGN) to attempt
to explain many of the observed properties of galaxies.  Specifically,
AGN feedback has been implicated as the primary cause behind
the differences between observed and theoretical galaxy luminosity
functions, the observed color bimodality of galaxies
\citep{white91,springel00,benson03,granato04,kauffmann04,springel05a} in
which galaxies are divided into distinct populations of blue
star-forming galaxies and red dead galaxies \citep[e.g.][]{strateva01}
with a pronounced deficit of galaxies with intermediate color, and the
very tight correlations observed between the masses of supermassive
black holes and their host galaxy spheroid's velocity dispersion
\citep{ferrarese00,gebhardt00} and luminosity \citep{marconi03}.  The
relative lack of galaxies with intermediate star formation rates
suggests that star formation ceases abruptly rather than gradually
\citep{bell04}, while the tight correlations of black hole masses with
the properties of their host bulges are taken as evidence that the two
are strongly co-eval.  In the various proposed feedback scenarios, the
supermassive black holes grow primarily by accretion until they become
sufficiently massive and energetic that thermal and/or radiative
feedback from their activity either heats the surrounding interstellar
gas or mechanically pushes it out of the host galaxy
\citep{crenshaw03,begelman04}.  This simultaneously deprives the host
galaxy of raw material from which to form new stars
\citep{springel05b,bower06,croton06} and starves the black hole,
truncating its growth \citep{silk98,fabian99}.

Despite the considerable utility of AGN feedback, there is scant
observational data to inform us about how or even if AGN feedback
operates in actual galaxies.  Arguably some of the most promising cases
are instances of mechanical feedback, such as the outflows that have
been detected as blue-shifted UV and X-ray intrinsic absorbers in
$\sim$50\% of Seyfert 1 galaxies and quasars
\citep{crenshaw99,crenshaw03,dai08}.  While it is clear that nuclear
outflows are common, it is still difficult to determine the mechanical
luminosity (critical for determining if the outflow can unbind
circumnuclear gas) and mass outflow rate to determine their impact.  A
good example of this difficulty is the work of \citet{krongold07} on the
time evolution of the ionization state of the X-ray absorbers relative
to the X-ray continuum in the warm absorber of NGC\,4051.  From the
absorption line variability they obtained good constraints on the
density and location of the absorbers from the black hole. Assuming a
biconical geometry, they measured a low outflow velocity with respect to
the escape speed from the black hole, and a corresponding low mass
outflow rate with respect to the accretion rate of the black hole. From
these values, they conclude that while the outflows might disrupt the
hot ISM, they are not capable of ejecting large amounts of interstellar
gas from the host. \citet{whittle02, whittle05, whittle08} reached a
similar conclusion for the nearby Seyfert galaxy Markarian\,78 using
visible-wavelength and radio data to measure the kinematics and
ionization state of extended emission-line regions surrounding the
active nucleus. They concluded that the biconical outflow in Mrk\,78 is
weak, slow and heavy, and hence insufficiently energetic to
significantly heat or unbind surrounding host galaxy gas.  

The results for these two galaxies present an apparent problem for AGN
feedback: in both cases the measured AGN outflows are too weak to
thermally or mechanically disrupt star formation in the host galaxy.
Nevertheless, these are just two objects, and analyses of other AGNs are
critical for further understanding the role of AGN feedback in galaxy
evolution.  Seyfert galaxies are of special interest for detailed
studies of AGN feedback as many are luminous enough to be accreting near
their maximal Eddington rate, yet close enough to study at relatively
high angular resolution (scales of 10--100\,pc for nearby examples)
using the {\it Hubble Space Telescope} and ground-based adaptive optics.

A very promising candidate for detailed study is the nearby Seyfert
galaxy Markarian\,573, which is well-known for its extended, richly
structured circumnuclear emission-line regions. Mrk\,573 features a prominent
ionization bicone and bright arcs and knots of emission-line gas
\citep{ferruit99,quillen99} that are strongly aligned and interacting 
with a kiloparsec-scale low-power radio outflow
\citep{pogge93,falcke98,ferruit99}. The association between the radio
lobes and the emission regions was studied in detail by
\citet{falcke98}, who proposed that the arcs result from gas cooling
after passing through a radio-induced radiative bow shock.  In contrast,
\citet{quillen99} argued that the arcs are instead a morphological
artifact of dust lanes in the galaxy being illuminated by the nuclear
ionizing continuum.  This is supported by visible and near-infrared
\hst\ images that show the emission-line arcs as extensions of
larger circumnuclear dust lanes that are illuminated when they pass into
the ionization cone.  Circumnuclear dust lanes like these are commonly
found in active and inactive galaxies alike \citep{martini03}, and in
the specific case of Mrk\,573 they appear to be shaped by the presence
of a nuclear bar \citep{martini01}, not unlike what is seen in galaxies
with inner bars regardless of the presence of nuclear activity.

\citet{ferruit99} examined several possibilities for the origin of the arcs
and extended emission-line regions using a combination of \hst\ images
and ground-based, integral-field spectroscopy.  They modeled the
excitation in the arcs as either shock features, linked to radio jets,
or pre-existing structures photoionized by the nucleus.  Using
standard emission-line diagnostics, they showed that the inner arcs are
excited by the central continuum source rather than by fast,
photoionizing shocks, citing as evidence that they detected no signs of a
strong kinematic interaction through a radio-induced shock. The
excitation of the outer arcs was not completely explained by nuclear
photoionization; these arcs require an external source of photons in addition
to the nucleus to account for the excitation levels.  However, their
analysis suggested that both the inner and outer arcs show little
evidence that they are shock excited.

In this paper we present new long-slit spectrophotometry of Mrk\,573
obtained with the Hubble Space Telescope Imaging Spectrograph (STIS).
The high angular resolution of STIS, combined with good velocity
resolution in the bright H$\alpha$+[\ion{N}{2}] emission lines allows us
to examine the excitation and kinematics in these regions in greater
detail than possible in previous studies.  Our analysis of Mrk\,573
follows a threefold approach.  First, we examine the kinematics of the
circumnuclear emission-line regions to separate kinematically-disturbed
gas from quiescent gas in the rotating disk of the galaxy (see sec.\,\ref{sec:los_vel}).  The
existence of these two kinematical components was clear from the earlier
ground-based work, but the greater angular resolution provided by {\it
Hubble} lets us pinpoint the regions of disturbed and undisturbed gas on
a scale of 10s of parsecs.  Next we measure the densities and
temperatures in the emission-line regions using standard nebular
diagnostic emission lines. We examine the excitation state of these
regions, comparing them with the predictions of published
photoionization and shock excitation models to clarify the nature of the
ionization mechanism in the bright emission line areas (see sec.\,\ref{sec:spectrophot}). Finally, we
undertake a quantitative analysis of the outflow energetics of Mrk\,573
using techniques developed by \citet{whittle08} to examine the degree of
feedback on the host galaxy surroundings (see sec.\,\ref{sect:whittle_quant}).  Throughout this paper we
adopt a distance of 74\,Mpc for Mrk\,573 ($v_{helio}=5150\pm11$\,\kms,
$H_0=70$\,\kms\,Mpc$^{-1}$), giving us a projected linear scale of
$\sim$360\,pc\,arcsec$^{-1}$. For reference, this is approximately half
the distance of Mrk\,78 ($v_{helio}=11137$\,\kms).

\section{Observations and Data Reduction}

\subsection{Hubble Space Telescope Observations}

\begin{figure}
\centering
\includegraphics[width=5in]{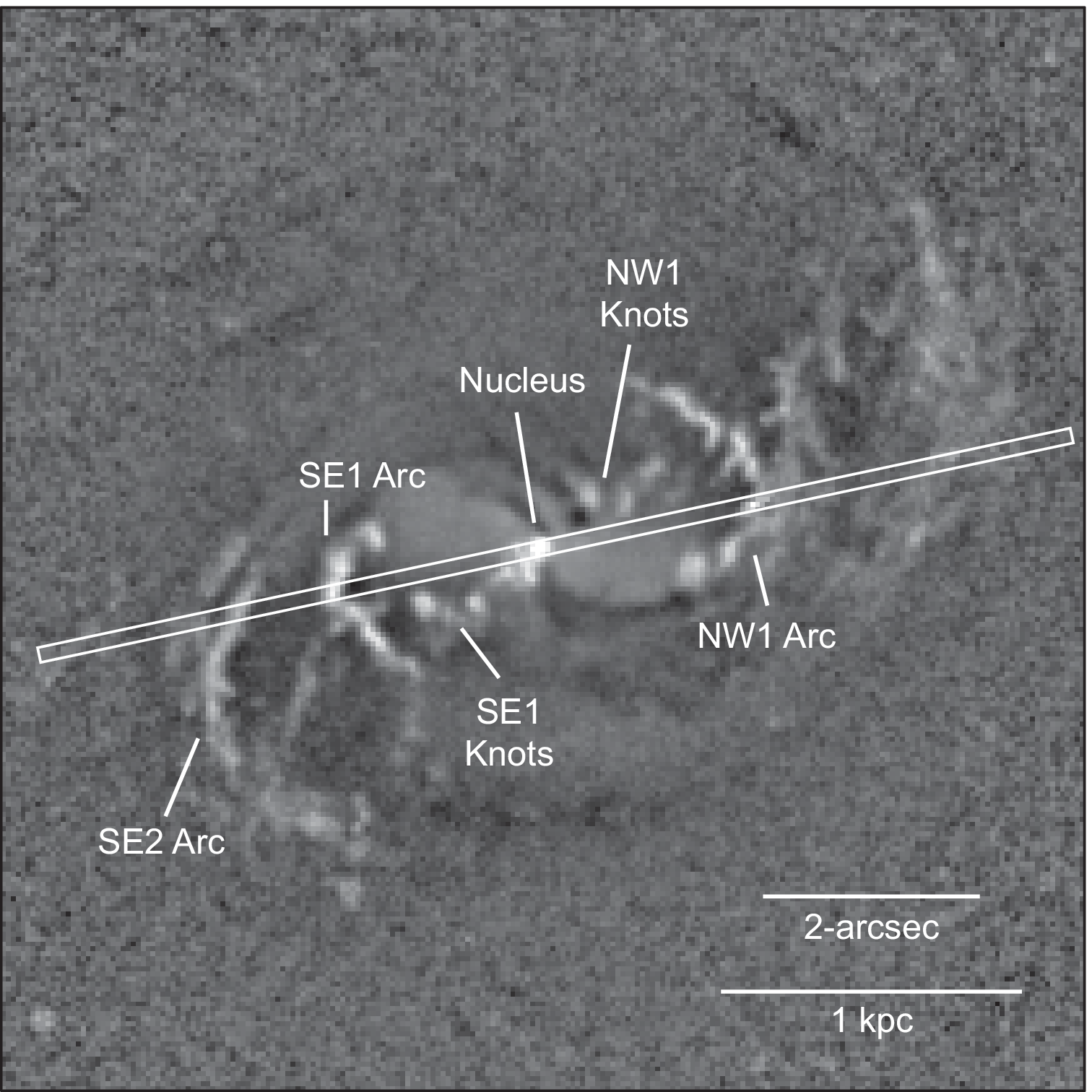}
\figcaption{
Contrast-enhanced structure map of the central 10\arcsec\ of Mrk\,573,
constructed from an archival WFPC2 F606W filter image by
\citet{pogge02}.  Emission-line regions appear bright and dust lanes
appear dark.  The STIS slit used for the long-slit spectra shown in
Figure~\ref{fig:lspec} is superimposed (the actual long slit extends
54\arcsec, but we show only the inner 10\arcsec).  The main
emission-line regions are labeled with the names adopted from
\citet{ferruit99}.  The image is oriented North up, East to the left,
with scale bars indicating the angular and approximate linear scales
in arcseconds and kiloparsecs, respectively.
\label{fig:em_regions}}
\end{figure}

We acquired spectra of the nucleus of Mrk\,573 on UTC 2001 October 17 using 
the {\it Hubble Space Telescope} and STIS, with the 52$\times$0.2 aperture.
Our data were obtained in one observation period lasting two orbits with two
spectral settings: the medium-dispersion G750M grating centered
on the H$\alpha$ emission line, ranging from
$\lambda\lambda6300-6850$\AA\, and the low-dispersion G430L grating
providing coverage from $\lambda\lambda2900-5700$\AA.  These configurations
provide FWHM spectral resolutions for extended sources of 2.2 and 10.9\AA\,
respectively, and angular resolution of 0\farcs051 pixel$^{-1}$.  The
spacecraft roll angle was unconstrained in our observation planning as
the data were acquired as part of a program to observe the nucleus
proper, and an attempt to constrain the slit position angle to the
apparent jet axis resulted in plan windows too short for practical
scheduling.  By pure good fortune, the slit position angle was
$-70.78$\degr\ , which is within $16$\degr\ of the radio-axis position
angle of $-54$\degr but still passes through the bright part of the
inner arcs, well within the ionization cone and the main
region of the nuclear outflow seen by
others. Figure~\ref{fig:em_regions} shows the location of the slit
relative to the emission-line regions on a contrast-enhanced ``structure
map'' created from an \hst\ F606W image \citep[see][]{pogge02}. 

The G750M spectra were acquired as a sequence of three exposures with
target exposure times of 1080, 1080, and 840s. The G430L observations 
consisted of two exposures of 805 and 840s duration. To avoid problems with hot pixels on the
STIS CCD, the observations were dithered by $\pm$0\farcs25 (5 pixels)
along the slit relative to the first spectrum. Wavelength calibration
lamp spectra were taken during Earth occultation.

\subsection{Data Reduction}

Our data were acquired after the primary (Side-1) STIS electronics
failed on UTC 2001 May 16.  Since the Side-2 electronics did not provide
closed-loop temperature control of the STIS CCD, the CCD temperature
varies with the ambient temperature of the spacecraft when the
thermoelectric cooler is run at a constant rate \citep{STIS_handbook}.
The result is that the dark rate varies with temperature, and the
standard dark calibration images used by the {\tt calstis} data
reduction pipeline are often a poor match to the actual dark rates in
the hot pixels.  To correct for this, we broke out of the {\tt calstis}
pipeline after the {\tt BASIC2D} step and corrected the dark pixels
individually before re-inserting them into the pipeline for the final
wavelength calibration, flux calibration, and geometric rectification
steps.  In brief, a custom program scanned the Data Quality Frame (DQF)
for each individual image and created a list of hot pixels.  This list
was used to fix all pixels in the science frame that had either negative
data values or that deviated from the median within a 7$\times$1-pixel
box centered on that pixel by more than 2$\sigma$. For this last step,
the long-axis of the analysis box is oriented along the dispersion axis
of the detector.  Once this hot-pixel surgery was completed for the
three G750M observations, the second and third images in the set were
then aligned with the first and co-added using the {\tt calstis} {\tt
ocrreject} task to detect and remove cosmic rays.  Remaining cosmic rays
missed by this step were removed by hand using the median filter routine
{\tt TVZAP} in {\tt XVista}.  The final combined, cleaned, spectral
image was then fed back into the {\tt calstis} pipeline to produce the
final calibrated long-slit spectrum.  An analogous process was used for
each of the two G430L spectra to create the final G430L long-slit
spectrum, but because only two images were available, more hand cleaning
of residual cosmic rays was required.  This process yielded superior
results by comparison with the default reduction pipeline. An analogous
procedure is described in \citet{rice06}.

\begin{figure}
\centering
\includegraphics[width=6in]{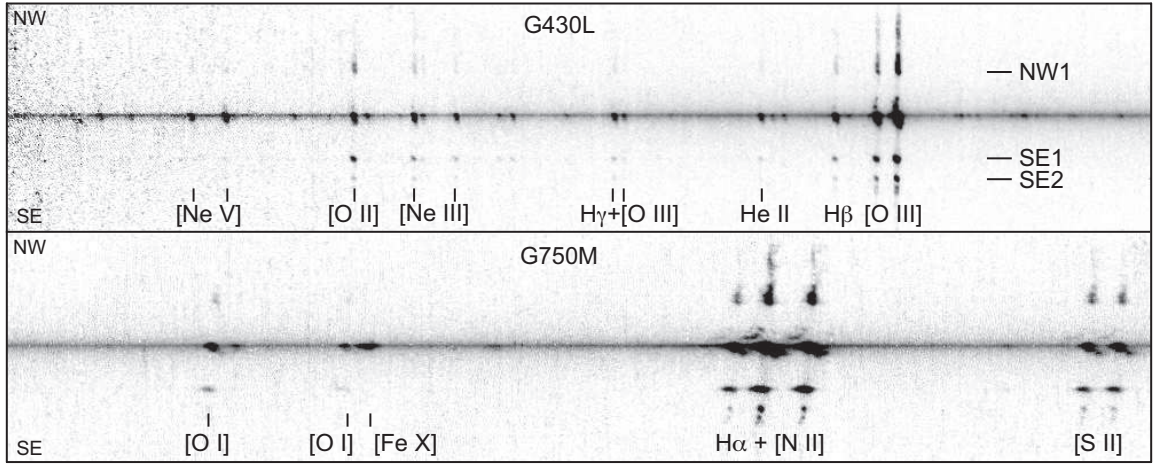}
\figcaption{
Long-slit 2D STIS spectra of Mrk\,573 taken with the G430L {\it (top)}
and G750M ({\it bottom}) gratings.  The central 10\arcsec\ of the STIS
slit is shown, with NW at the top and SE at the bottom of each panel.
The G430L spectrum extends from $\lambda2900$\AA\ to $\lambda5700$\AA,
while the G750M spectrum extends from $\lambda6300$\AA\, to
$\lambda6850$\AA.  Prominent emission lines are labeled.
\label{fig:lspec}}
\end{figure}

An additional consequence of using the STIS Side-2 electronics is that
our long-slit spectra are affected by low-amplitude (2--3\,ADU
peak-to-peak) fixed-pattern noise with a $\sim$3-pixel horizontal scale
that takes the form of a regular herring-bone noise pattern when viewed
at high contrast. Attempts to remove this component using fast
fourier-transform filtering only made things worse.  Its primary effect
is to increase the effective readout noise of the device by about
1\,$e^{-}$\,pix$^{-1}$ for the gain=1 mode of our data.  When extracting
faint outer regions of the long-slit spectra, we took particular care to
be wary of this fixed-pattern component in interpreting our spectra.
The final, calibrated long-slit spectra are shown in
Figure\,\ref{fig:lspec}.  We have adopted the nomenclature of
\citet{ferruit99} for the specific emission-line regions we examine below.

\subsection{Spectral Extraction and Line Measurements}
\label{sec:extract}
The extended emission-line region of Mrk\,573 consists of a series of
arcs and knots located northwest and southeast of the nucleus along the
general direction of the radio outflow.  Our STIS slit intersects the
arcs but just misses a group of bright knots located along the main
radio axis as shown in Figure\,\ref{fig:em_regions}).  We shall focus
our attention on four specific regions: the SE1 arc, SE2 arc, the
nucleus, and the NW1 arc \citep{ferruit99}. The arcs at regions SE1 and
NW1 are located at a projected angular distance of $\sim$2\arcsec\ on
either side of the nucleus, and the SE2 arc is $\sim$3\arcsec\ from the
nucleus. There is another arc in the Northwest region that is roughly
symmetric with the SE2 arc, but the emission lines from this region are too
faint and diffuse for detailed analysis with our spectra.

We extracted two 1D spectral data sets from the 2D spectra.  The first
set is integrated spectra of specific regions of interest for measuring
the density, temperature, and excitation state of the line-emitting gas.
For these we first created slit intensity profiles for the
H$\alpha$+[\ion{N}{2}] emission lines in the G750M spectrum and for
[\ion{O}{3}]\,$\lambda$5007\AA\ in the G430L spectrum, and for adjacent
line-free continuum regions bracketing these lines.  The derived mean
continuum profiles were then subtracted from the line profiles to
produce pure emission-line profiles for each line.  These profiles were then used 
in conjunction with the 2D spectra of the nuclear regions (see Figure\,\ref{fig:lspec}) 
to precisely locate the windows for extracting 1D spectra of the regions we wished to examine.

The second set of spectra were extracted from the G750M data for making
kinematic measurements - line centroids and velocity widths - using the
H$\alpha$ and [\ion{N}{2}] emission lines.  For these we extracted 1D
spectra in contiguous increments along the slit moving radially outwards
from the nucleus.  In bright regions we extracted spectra from
single-pixel apertures (0\farcs2$\times$0\farcs05), and as the regions
became fainter at larger radii, we increased the width of the extraction
window up to a limit of 4 pixels wide (0\farcs2$\times$0\farcs2) to
improve the signal-to-noise ratio in the lines.  This gives us nearly
continuous spatial sampling along the slit for measuring line-of-sight
velocities and emission-line velocity dispersions.

After identifying individual emission lines, we measured the line
parameters with the LINER interactive line analysis program developed at
Ohio State.  For each spectral line, we define a local continuum by
averaging over adjacent line-free regions and then fit the lines with a
single- or multi-component Gaussian to resolve blends or multiple
velocity components if present.  We derive the line centroid, Full-Width
at Half Maximum (FWHM), and integrated line intensity for each
emission-line component present.  When fitting doublet lines of a single
ionic species (e.g., [\ion{N}{2}]\,$\lambda\lambda$6548,83\AA\ and
[\ion{S}{2}]\,$\lambda\lambda$6716,31\AA) we fit both lines together,
imposing physically-motivated constraints on the profile parameters to
improve the fidelity of the fits.  For the [\ion{S}{2}] doublet, we
constrain the relative line centroids in velocity space and match them
in line width, but leave the relative intensities (which are correlated
with density) unconstrained.  For the [\ion{N}{2}] doublet lines, which
arise out of the same upper excited level, we constrain their relative
line centroids and widths and further constrain their relative
line strengths as dictated by the ratio of their radiative transition probabilities.
This greatly improves the quality of the fits when deblending these
lines from the H$\alpha$ emission line that lies between them.  Some of
the emission lines appear double-peaked, particularly in the kinematic
spectral extractions.  In these cases we model the two peaks as
Gaussians and apply similar rules for setting constraints on the
physically related line pairs.  Finally, if the line of interest is very
faint but clearly detectable and unblended, we extract the line
intensity by direct integration without attempting to fit a line
profile.  The RMS uncertainty in the continuum fits and a simple model
for the pixel-to-pixel signal to noise ratio is used to estimate the
overall quality of the fit and the uncertainties on the best-fit line
parameters.

We make additional checks of our fit-derived line parameters in a number
of ways.  The total line intensities are compared to a numerical
integration of the blend itself without fitting to a particular line
shape.  Line FWHMs from the fits are compared to direct measurements of
the FWHM using profile tracing algorithms that assume no underlying
profile shape (this works best for the brightest emission lines with the
least blending and highest signal-to-noise ratio).  Finally, the
Gaussian fit derived line centroids for well-isolated lines are compared
to intensity-weighted centroids derived from the central 5 pixels about
the peak.  All of these give us further estimates of the relative
uncertainties, and help alert us to possible systematics that would be
missed by simply examining the RMS residuals.

The measured fluxes for various emission lines are listed in
Table\,\ref{tab:regional_feat_fluxes}.  In general, line centroids and
FWHM are measured to $\pm$0.1\AA\, or $\sim$20\,\kms\ for the H$\alpha$
and [\ion{N}{2}] emission lines, with some degradation in fainter
regions.  Measurement uncertainties for total line fluxes are typically 
$\pm$10\% in the brighter lines, increasing to $\pm$20\% in the faintest 
regions measured.

\section{Spectral Analysis} \label{sec:specanalysis}

We pursue two different analyses of our spectra.  The first uses the
kinematic spectra (sec. \ref{sec:los_vel}) to derive outflow speeds 
and separate outflowing gas from ambient
ISM gas in orderly rotation in the disk.  Combining the outflow
kinematics with the emission-line geometry seen in the direct images
provides us with constraints on the configuration of the outflow, allowing us to 
deproject the observed radial velocities and thus determine the actual 
physical outflow velocities. 

The second line of analysis uses emission line diagnostics from the
spectrophotometric set of spectra (sec. \ref{sec:spectrophot}) to estimate the
density, temperature, and excitation state of the gas in these regions.
These are compared to published models of shock- and photo-ionized gas
to determine the excitation mechanism in these regions.

\subsection{Emission Line Kinematics}\label{sec:los_vel}

\begin{figure}
\centering
\includegraphics[width=6.3in]{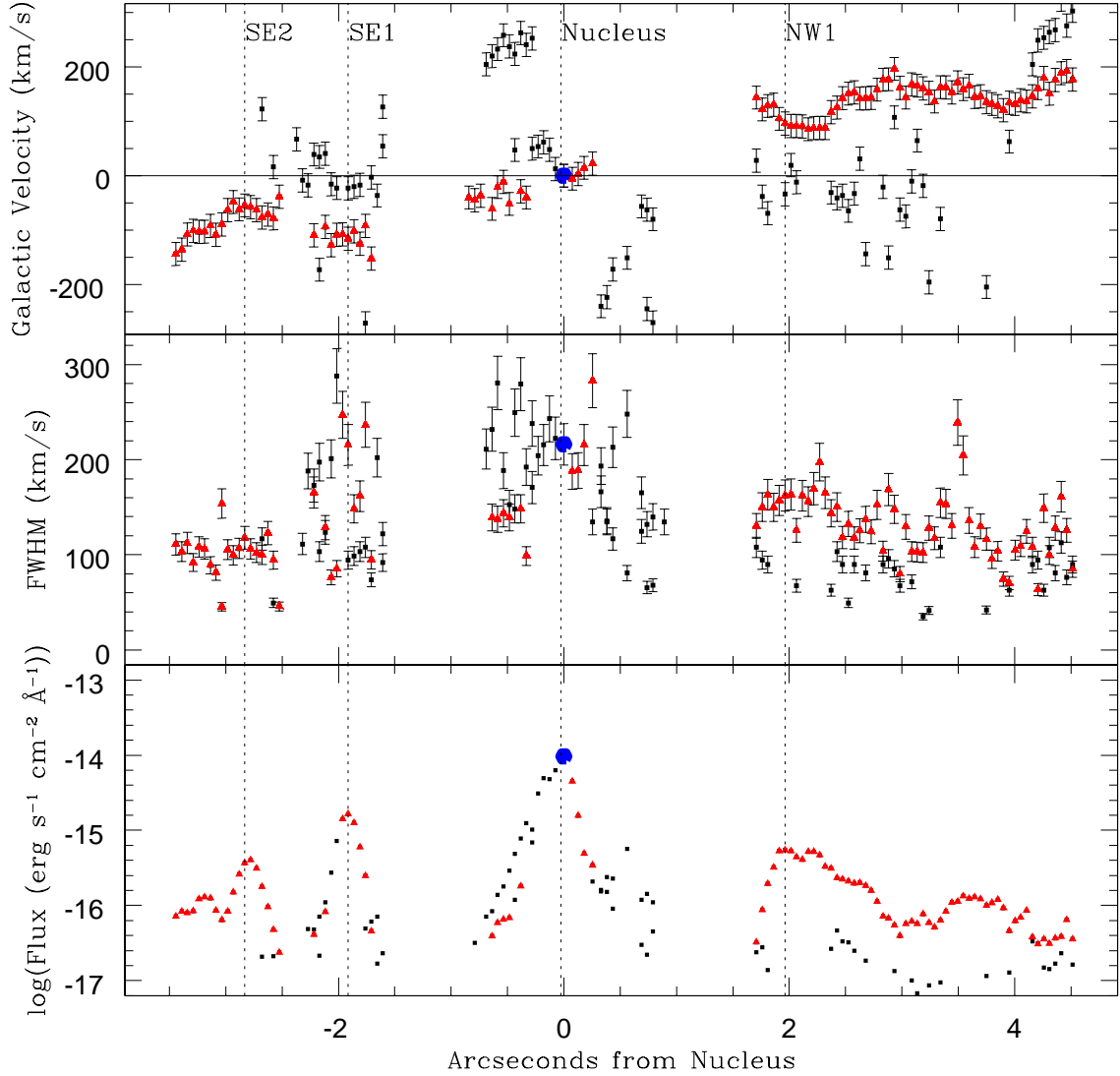}
\figcaption{
The H$\alpha$ radial velocity profile ({\it top}), FWHM profile ({\it
middle}), and total line intensity profile ({\it bottom}) along our slit
for Mrk\,573 as a function of distance from the nucleus. The red
triangles represent rotating disk component, the black squares
correspond to the biconical outflow region, and the large blue filled
circle represents the nucleus.  The centroids of the individual emission 
line regions are marked by the dotted vertical lines.\label{fig:halpha_rotcurve}}
\end{figure}

\begin{figure}
\centering
\includegraphics[width=6.3in]{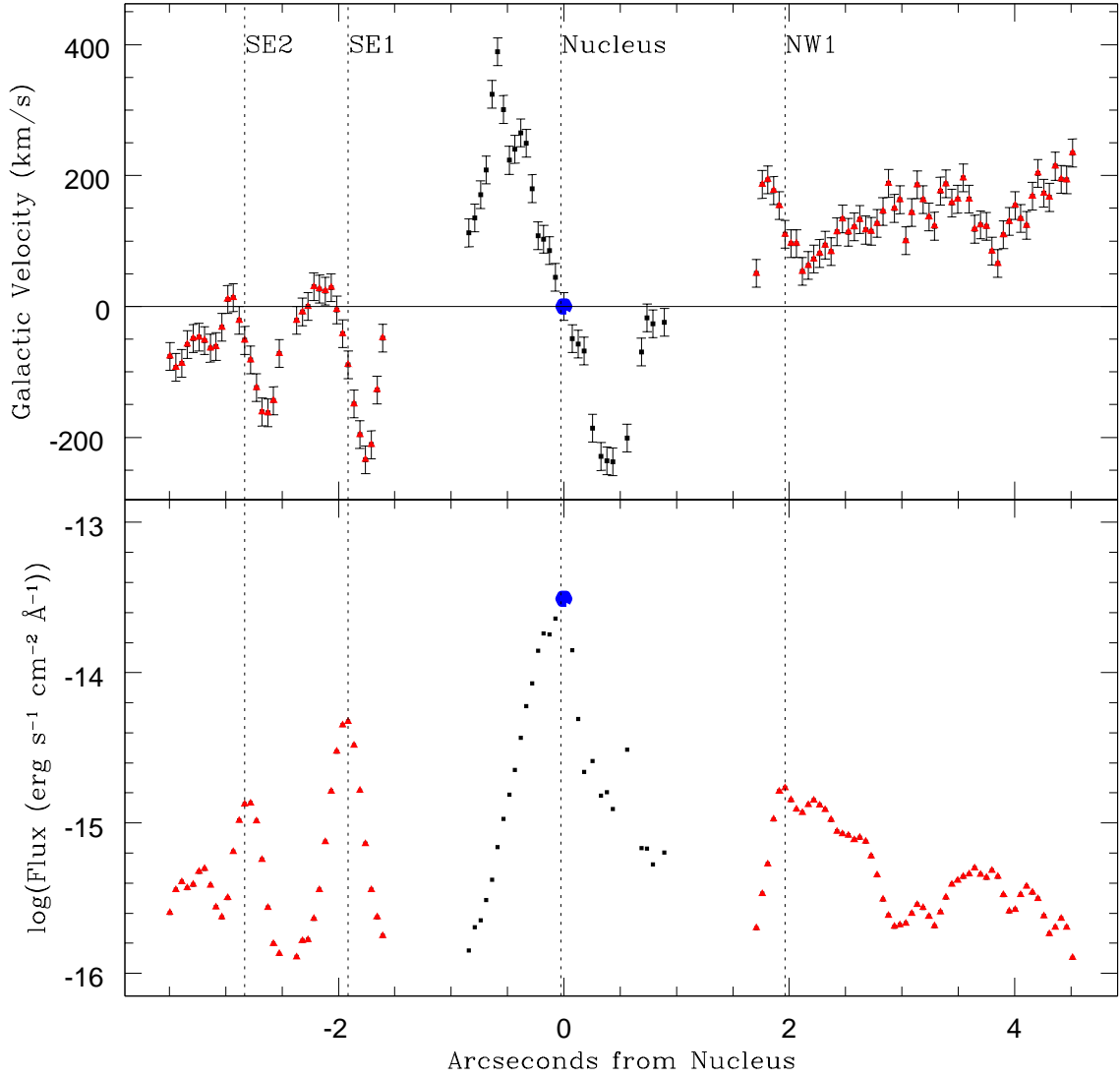}
\figcaption{
Same as Figure~\ref{fig:halpha_rotcurve} for the 
[\ion{O}{3}]\,$\lambda$5007\AA\ emission line. The FWHM curve is not shown as 
the resolution of the G430L grating is low dispersion, resulting in broader features. \label{fig:oiii5007_rotcurve}}
\end{figure}

Figures \ref{fig:halpha_rotcurve} and \ref{fig:oiii5007_rotcurve} show
the radial velocities along our long slit for the H$\alpha$
$\lambda$6563\AA\ and [\ion{O}{3}] $\lambda$5007\AA\ emission
lines\footnote{Radial velocity measurements for H$\beta$ $\lambda$4861,
[\ion{N}{2}] $\lambda\lambda 6548, 6583$, and [\ion{O}{3}]
$\lambda\lambda$4363, 4959\AA\ lines show consistent results and are available upon request.}.  A
number of distinct kinematic systems are present.  The largest-scale
pattern is a rotating disk extending out as far as we can trace ionized
gas (roughly $\pm$4\arcsec), with the NW side receding and the SE side
approaching us and an amplitude of roughly $\pm$125\,\kms.  In the inner
1\arcsec\ on either side of the nucleus the kinematics are dominated by
a coherent system of double-valued radial velocities (separated by nearly
300\,\kms) with a red/blue splitting pattern characteristic of a
biconical outflow \citep[e.g.][]{crenshaw00,das05}.  Finally, in the
inner 0\farcs5 around the nucleus we see various additional H$\alpha$
velocity components with no obvious pattern associated with the inner
parts of the outflow region.  The high-velocity components of the
biconical outflow pattern and the large-scale rotation pattern are also
visible in the [\ion{O}{3}]\,$\lambda$5007\AA\ line-of-sight velocities
(Figure\,\ref{fig:oiii5007_rotcurve}) despite the lower dispersion of
the G430L grating.

\begin{figure}
\centering
\includegraphics[width=5in]{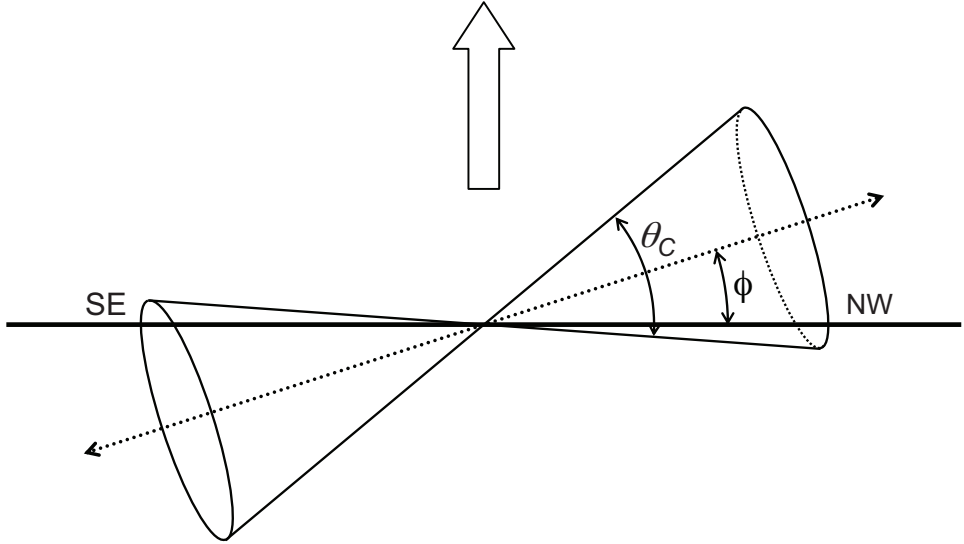}
\figcaption{
Schematic of the inferred outflow geometry viewed perpendicular to the
line of sight towards Earth (indicated by the large arrow).  This view
is from the south looking towards the nucleus.  The ionization cone is
shown with the 45\degr\ opening angle ($\theta_C$) measured by
\citet{wilson94} and is tilted by angle 18\degr\ ($\phi$) with respect 
to the plane of the sky, as estimated from
our kinematics (see section\,\ref{sec:los_vel}).  In this geometry, the NW 
outflow is emerging out of the plane of the sky towards us, while the SE 
flow is going into the sky.
\label{fig:geometry}}
\end{figure}

In the biconical outflow models of \citet{crenshaw00} and \citet{das05}
emission from the front of the cone (the surface closest to the
observer) is blueshifted and the back of the cone is redshifted.  When the
bicone axis is at high inclination angles (the cone axis near the line
of sight) there is a strong blue-to-red asymmetry in the radial
velocities, while at low inclinations (the axis near the sky plane), the
radial velocity pattern is more symmetric \citep{das05}.  In the context
of these models, our radial velocity maps indicate that the biconical
outflow in Mrk\,573 has a relatively low inclination with the NW cone
directed toward us.  Using the velocities from our H$\alpha$ emission
line, the observed ionization cone opening angle of $\theta_C=45^o\pm10$
\citep{wilson94}, and assuming a constant velocity, $v_{f}$, of material 
in a shell along the cone and a slit roughly aligned with the cone axis,
the \citet{das05} outflow model gives projected velocities of
\begin{eqnarray}
v_{Red}&=&v_{f}\sin(\theta_C+\phi)\approx200 \textrm{km/s} \\
v_{Blue}&=&v_{f}\sin(\theta_C-\phi)\approx100 \textrm{km/s}
\end{eqnarray}
for the red and blue sides of the cone, respectively.  The cone opening
angle ($\theta_C$) is related to these velocites and the cone
inclination angle ($\phi$) by
\begin{equation}
\tan(\theta_C)=\Big(\frac{v_{Red}+v_{Blue}}{v_{Red}-v_{Blue}}\Big)\tan(\phi)\approx3\tan(\phi) 
\end{equation}
which gives a bicone inclination angle of $\phi=18^{o}\pm4^{o}$ and 
$v_{f} = 220$ km s$^{-1}$ for the nuclear outflow in Mrk\,573. We model 
this outflow geometry in Figure\,\ref{fig:geometry}. 

Between about 1--1\farcs5 from the nucleus there is very little
emission line gas, and so we have no measurements in these regions.
This region between the nucleus and the two close emitting arcs 
likely represents the nearly evacuated ``bubbles'' resulting 
from the biconical outflow.  Beyond about 1\farcs5 on either side of
the nucleus a coherent pattern of velocities consistent with a rotating
disk of relatively undisturbed gas extends out to the limits of our
long-slit data.  This sense of rotation (SE approaching, NW receding)
is opposite the direction of the biconical outflow which helps to
distinguish these components.  At the location of the bright arcs (SE1
and NW1), the kinematics are disturbed: the radial velocities
split by $\sim$100\,\kms relative to the rotating disk pattern, and the
velocity dispersion jumps dramatically in the SE1 arc to nearly 300\,\kms.
This can be seen in the 1D radial velocity profile
(Figure\,\ref{fig:halpha_rotcurve}, middle panel) and in the 2D spectrum
(Figure\,\ref{fig:lspec}) where the H$\alpha$, [\ion{N}{2}] and
[\ion{S}{2}] emission lines are all suddenly broader.  The effect is
less obvious in the fainter NW1 arc, although in the 2D spectrum there
is faint but noticeable broadening of the H$\alpha$ and
[\ion{N}{2}]$\lambda$6583\AA\ lines.  These are the regions that
\citet{ferruit99} identified with the working surface between the radio 
outflow and ambient gas in the galaxy.  The SE2 arc appears to have similar 
kinematic disturbance as SE1 and NW1 in the [\ion{O}{3}] velocity map; however, 
it is not consistently double-valued in H$\alpha$, like the other two arcs (see 
Figure\,\ref{fig:halpha_rotcurve},\,\ref{fig:oiii5007_rotcurve}). This 
implies that SE2 is not significantly kinematically disturbed as SE1 and NW1 are. 
Throughout this analysis, we include calculations for SE2, but our primary focus 
is on the effect of an outflow as manifested in SE1 and NW1. 

\subsection{Spectrophotometry} \label{sec:spectrophot}

\begin{figure}
\centering
\includegraphics[width=5in]{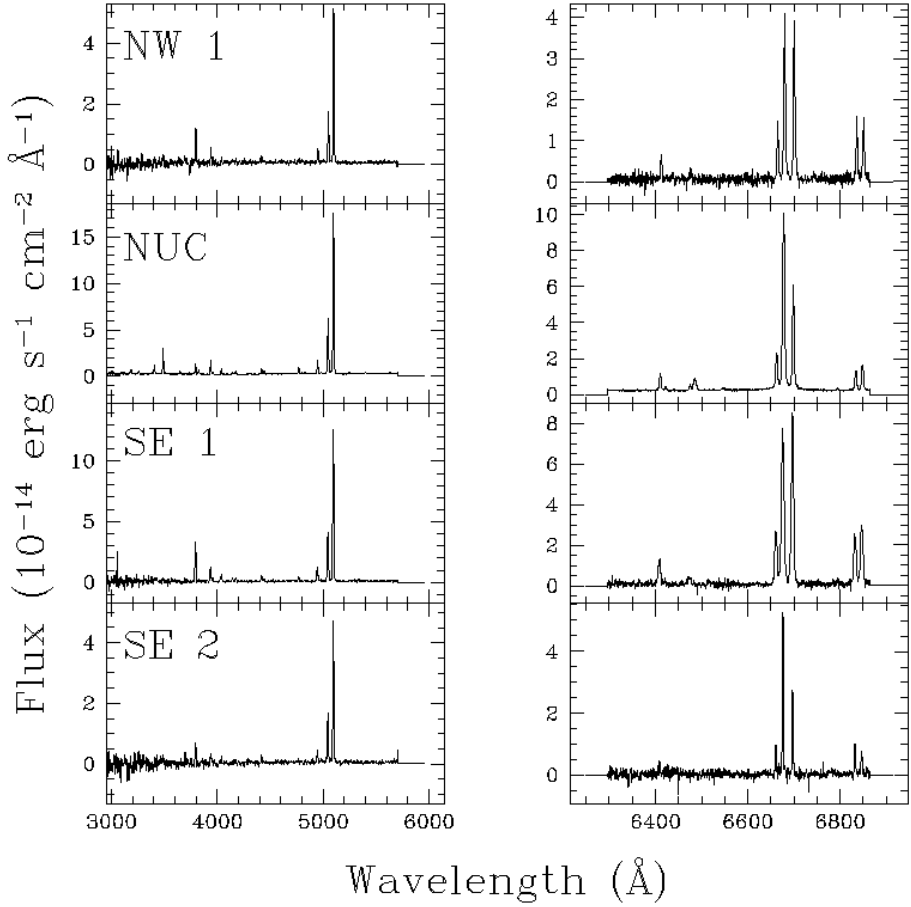}
\figcaption{
Extracted spectra of emission-line regions in Mrk\,573. The left panels contain 
the G430L spectra and are labeled with the associated region from 
Figure~\ref{fig:em_regions}. The right panels contains the G750M spectra. 
Each extraction region was chosen to include all of the emission of each 
region; the angular extent of these regions is listed in Table\,\ref{tab:regional_feat_fluxes}. 
All of the regions have [\ion{O}{3}] $\lambda\lambda$\,4363, 4959, and 
5007\AA\ and [\ion{S}{2}]$\lambda\lambda$\,6716 and 6731\AA\ emission features, 
which were used to determine the density and temperature. The nucleus includes 
additional high-ionization features, such as [\ion{Fe}{10}] $\lambda$6375\AA\, 
which indicates the presence of close-in material at higher temperatures.  
\label{fig:regions}}
\end{figure}

The spectra of the nucleus and bright, off-nuclear regions are
remarkably rich in emission lines, as can be seen in
Figure\,\ref{fig:regions} where we plot the spectra of the nucleus and
the brightest off-nuclear regions.  Of particular interest are the lines of
H$\alpha$, H$\beta$, [\ion{N}{2}]$\lambda\lambda$6548,6583\AA,
[\ion{O}{3}]$\lambda\lambda$4363,4959,5007\AA, and
[\ion{S}{2}]$\lambda\lambda$6716,31\AA\ which are used to estimate
internal extinction, densities, and temperatures in the gas, and provide
diagnostics of shock versus photoionization.  In
addition, we can observe a number of fainter high-excitation lines of
interest, including [\ion{Fe}{10}]$\lambda$6375\AA\ and the
[\ion{Ne}{5}]$\lambda$3426\AA\ line.

\subsubsection{Densities and Temperatures}
\label{sec:region_dt}

The densities of the emission-line gas are estimated using the
[\ion{S}{2}]\,$\lambda\lambda$6716,31\AA\ doublet ratio and the
theoretical calculations of \citet{cai93}. The gas temperatures are
calculated using the [\ion{O}{3}]\,$\lambda\lambda$4363,4959, and
5007\AA\ emission lines following the method described by
\citet{osterbrock89}.  Because the [\ion{O}{3}] emission-line ratio
$I(4959)+I(5007)/I(4363)$ is reddening-sensitive, we estimate the effects 
of extinction by assuming a simple screen geometry and using the \ion{H}{1}
Balmer decrement assuming Case-B recombination.  The small
differences between the observed and emitted [\ion{O}{3}] line ratios in
Table\,\ref{tab:ne_temp_fluxes} indicate that reddening in these 
regions is negligible. 

Because the [\ion{S}{2}] doublet density method is weakly temperature
dependent and the [\ion{O}{3}] temperature estimate is weakly density
dependent \citep[see][]{osterbrock89}, we have adopted an iterative
approach, whereby we first estimate the [\ion{O}{3}] temperature in the
low-density limit, evaluate the [\ion{S}{2}] density at that
temperature, and then use that density to re-evaluate the [\ion{O}{3}]
temperature, iterating until the solutions converge to within the
measurement uncertainties.  The densities we find range from
760-2500\,cm$^{-3}$, and the temperatures range between
10000-14500\,K. The resulting temperature and density estimates for each
region and their uncertainties are given in
Table\,\ref{tab:ne_temp_fluxes}.

\subsubsection{Emission Measures}
\label{sect:emission_measure}

The arcs are high-surface brightness emission-line features, but the
``bubbles'' between SE 1, NW 1, and the nucleus (B$_{SE}$ and B$_{NW}$ 
respectively) are very low surface brightness. As the bubbles have 
only very weak emission
lines, we cannot estimate the temperature and density using standard
metal-line diagnostics. Instead, we make an order-of-magnitude estimate
of the mean density $\langle n_e\rangle$ using the emission measure
($EM$) derived from the surface brightness of H$\beta$ ($I(H\beta)$), an
estimate of the line-of-sight thickness of the bubbles, $D$, and the
volume filling factor $f$:
\begin{equation}
\label{eq:I}
EM = \frac{4\pi I(H\beta)}{h\nu_{H\beta} \alpha_{H\beta}^{eff}(T)} = 
\int_{los}n_{e}n_{p}ds \approx \langle n_{e}^{2} \rangle D f
\end{equation}
We assume that the temperature is roughly constant, and that the
bubbles are roughly spherical (i.e., as thick as they are wide) and have
unit filling factor.  A first estimate assumes a temperature of
15000\,K, slightly higher than the temperature of the nuclear region, to
evaluate the effective recombination coefficient, giving a mean density
of $\sim$1\,cm$^{-3}$ in the bubbles (see
Table\,\ref{tab:ne_temp_bubbles}).  At this temperature and density the
gas in the bubbles is out of thermal pressure balance with the gas in
the emission-line arcs just outside them by a factor of $\sim$500-1500.  The
bubbles may instead be more like supernova bubbles with internal
temperatures of $\sim10^6$\,K; adopting $T=10^6$\,K gives
emission-measure derived mean densities an order of magnitude larger
(see Table\,\ref{tab:ne_temp_bubbles}), bringing the bubbles into rough
thermal pressure balance.  The latter temperature is consistent with
thermalization of bulk motions with velocities as observed, i.e.
$\sim 200 - 300$ km s$^{-1}$.

\subsubsection{Nuclear Emission}
\label{sec:nuclearemission}

While the bubble regions have only very weak emission lines, the nuclear
emission spectrum has many high ionization lines that are undetected in
other emission-line regions, notably [\ion{Fe}{10}]$\lambda$6375\AA,
[\ion{Ar}{5}] $\lambda$6435\AA, and [\ion{Ar}{4}] $\lambda \lambda$4711,
4740\AA\ (see Figure\,\ref{fig:regions}). The presence of these 
lines indicates that the photons emitted from the nucleus have a 
sufficiently hard spectrum to substantially ionize the surrounding
material.  Some of these lines appear to be double peaked; in particular
[\ion{Fe}{10}] $\lambda$6375\AA\, has a broad, flat top. This feature is
best fit by two blended Gaussian lines, at $\lambda$6483 and
$\lambda$6486\AA, respectively, corresponding to line-of-sight
velocities of $-$46 and +92\,\kms. This line splitting likely represents
highly excited gas at the base of the biconical outflow.

\subsubsection{Emission Line Diagnostics} 
\label{sec:eldiagnostics}

Shock-ionized gas will have different relative emission line strengths
than gas photoionized by the active nucleus.  Collisionally excited UV
resonance lines provide the best discriminants between shock and
photoionization due to the higher temperatures expected in the shocked
gas \citep{dopita02,allen98}. Unfortunately, we do not have near-UV
spectra of Mrk\,573 and therefore we must use visible-wavelength
indicators.

To evaluate the importance of shock excitation for the arcs, we compare
the fluxes of our observed optical emission lines
(Table\,\ref{tab:regional_feat_fluxes}) to the predictions from models for
ionizing shocks with precursors, photons traveling ahead of the shock
front, and nuclear photoionization calculated by \citet{whittle05} and
\citet{allen08}.  The shock models cover a range of shock speeds
($v_s=200-1000$\,\kms) and magnetic field strengths, $B/\sqrt
n=0-10\mu$G\,cm$^{3/2}$.  We adopt two illustrative photoionization models from
\citet{whittle05}. The first is a power-law photoionization model for
optically thick gas with a density of $10^2$\,cm$^{-3}$ and a nuclear
continuum with a spectral index of $\alpha=1.4$ (where $f_\nu \propto 
\nu^{-\alpha}$).  This model covers
photon energies from 50\,keV to 100\,$\mu$m with an ionization parameter
range of ${\rm log} U=-4.0$ to $0.5$.  The second is the ``$U_{dust}$''
model that is similar to the first but includes a realistic dust component
that accommodates changes in the internal structure of dusty clouds in
response to the incident radiation field from \citet{groves04}.  This
second model is for a density of $10^3$\,cm$^{-3}$, continuum spectral
index of $\alpha=1.4$, and a metallicity twice the solar abundance.

\begin{figure}
\centering
\includegraphics[width=6in]{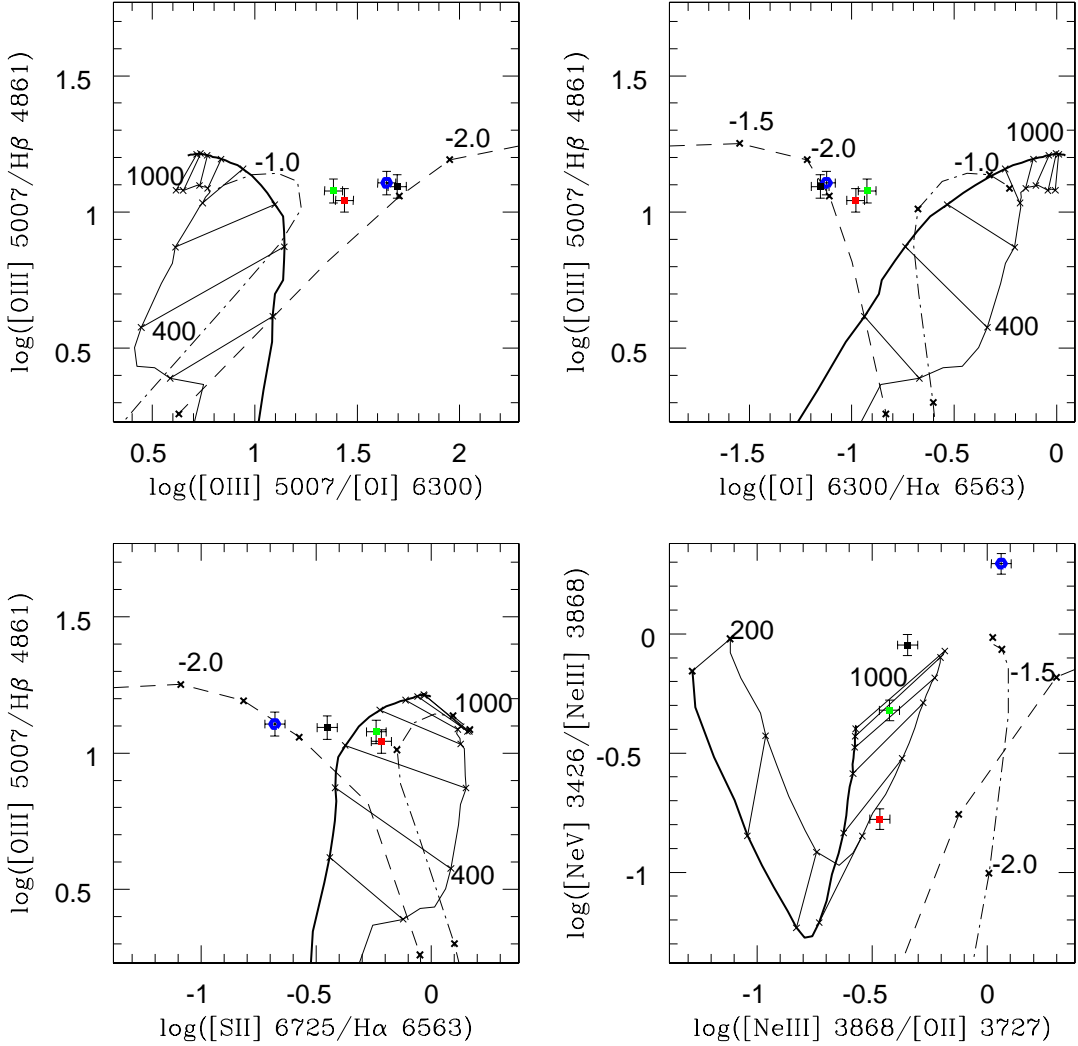}
\figcaption{
Optical ionization diagnostic diagrams. These diagnostics 
are the best at distinguishing between shocked and photoionized 
regions. Inconclusive ionization diagnostics are not included, but are 
available by request. The dashed line is U$_{1.4}$,
photoionization of an optically thick medium by power-law emission. The
crosses mark $\log U_{1.4}=-1.0, -1.5, -2.0, -2.5$ and $-$3.5. The
dot-dash line represents U$_{dust}$, power-law photoionized dusty
gas. Crosses mark $\log U_{dust}=-3.0,-2.0, -1.0$, and 0.0. The solid
lines represent shocks from \citet{allen08}. The thin line is a photoionizing
shock with B/$\sqrt n \approx 0\,\mu$G\,cm$^{-3/2}$. The thicker solid line connected to the thin
is a photoionizing shock with B/$\sqrt n = 10\,\mu$G\,cm$^{-3/2}$. These two shocks are
modeled for velocities of 200--1000\,km\,s$^{-1}$. The crosses mark every
additional 100\,km\,s$^{-1}$. These models are shocks and precursors and 
assume a pre-shock density of 1\,cm$^{-3}$ and solar abundance.  
The points represent each of the four
studied regions of Markarian 573. Blue represents the nucleus, green is
the NW1 region, and red and black are SE1 and SE2,
respectively.  
\label{fig:diagram1} }
\end{figure}

Figure~\ref{fig:diagram1} shows four emission-line diagnostic diagrams
that provide the cleanest discrimination between shock and central
photoionization for the lines available in our spectra. The upper two
panels show two diagnostics using the [\ion{O}{1}] $\lambda 6300$\AA\
line: [\ion{O}{3}]/H$\beta$ vs.\ [\ion{O}{3}]/[\ion{O}{1}] and
[\ion{O}{3}]/H$\beta$ vs.\ [\ion{O}{1}]/H$\alpha$.  In both cases the
data are in best agreement with the simple, $U_{1.4}$ photoionization
model and appear to disfavor the shock sequences and to a lesser extent
the $U_{dust}$ model.  However, these results should be viewed with some
caution as the [\ion{O}{1}] line is difficult to model: in
photoionization models the [\ion{O}{1}] is formed in a partially ionized
zone heated by X-ray photons, while shock models are known to
overpredict the [\ion{O}{1}] line strength by as much as an order of
magnitude \citep[e.g.][]{groves04}.

The lower left panel of Figure~\ref{fig:diagram1} shows the
[\ion{O}{3}]/H$\beta$ vs.\ [\ion{S}{2}]/H$\alpha$ diagnostic in which
the photoionization and shock models are essentially orthogonal to one
another.  Here the nucleus and SE2 region are in better agreement with
the photoionization model, while the NW1 and SE1 arc regions are
ambiguous between photoionization and a relatively fast shock
($v>500$\,\kms) with a strong magnetic field
($B/\sqrt{n}=10\,\mu$\,G\,cm$^{3/2}$), while they are in good agreement
with the $U_{dust}$ model. In the last panel we show the
[\ion{Ne}{5}]/[\ion{Ne}{3}] vs.\ [\ion{Ne}{3}]/[\ion{O}{2}]
diagnostic. These line ratio measurements are similar to the previous
panel, in particular the agreement between the NW1 and SE1 measurements
and the shock models, while the nucleus and SE2 region do not agree well
with any of the models. One caveat with this diagram is that
[\ion{Ne}{3}] may be overpredicted by the models
\citep{whittle05}.

While none of these line ratios are ideal discriminants between shock
and photoionization, we nevertheless see consistent trends in our
analysis that favor photoionization.  One interesting aspect of these
data is that the SE1 and NW1 arc regions, which trace the most
morphologically-obvious bowshock structures, are more consistent with
shock models than either the nucleus or the SE2 region. This also supports 
our assertion that SE2 is not strongly interacting with a biconical outflow. 
Additional far UV data could prove useful to further investigate the 
strengths of any shocks in these regions; however, based on these and 
previous arguments we conclude that the emission line ratios in all 
four regions are more consistent with photoionization than with shocks. 
While they themselves are not a clear cut indicator of photoionization, 
they do support the findings from kinematics and analysis of energetics.

\subsection{Discussion: Ionization Mechanisms}
\label{sec:iondisc}

Our analysis thus far adds additional detail to the picture for the
inner regions of Mrk\,573 developed by previous work: a hot biconical
outflow emerging from the nucleus that is running into the host galaxy
at the location of the bright emission-line arcs.  Our kinematic data
show that this outflow is mildly inclined with respect to the host
galaxy disk ($18\pm4^{o}$), and there is a modest kinematic disturbance at
the location of the arcs.  The inner part of the outflow between the
dense base of the bicone and the arcs has low density
($\sim1$\,cm$^{-3}$)

The nature of the arcs from previous work has been equivocal.  On the
one hand, \citet{ferruit99} inconclusively modeled these same emission
regions as either pre-existing structures or shock features, while on
the other \hst\ images show dust lanes that extend beyond the ionization
cones which led \citet{quillen99} to suggest that the arcs of Mrk\,573
do not result from bow shocks. Our analysis of the densities of the 
different emission line regions reveal that the arcs and nucleus have 
similar densities, implying that the arcs were not formed by a fast 
shock. A fast shock would have slammed into the surrounding material, 
making it more dense, which we do not observe. The density values 
support the morphological argument of \citet{quillen99} that the 
central photons are ionizing pre-existing dust lanes. In turn, this 
supports the suggestion of \citet{ferruit99} that any shock from the 
central AGN is too weak and slow to photoionize the surrounding material.

Analysis of the evacuated regions, the bubbles, compared to the emission line 
regions also supports central photoionization of dust lanes. If
these bubbles were rapidly expanding and compressing the surrounding
material, they would shock ionize the arcs. In this scenario we expect a
higher pressure inside the bubbles than the arcs to cause the
expansion. Our calculation of the emission measures (see
sec.\,\ref{sect:emission_measure}) reveals that the bubbles have
$\sim$500-1500 times less thermal pressure than the arcs, assuming a temperature 
of 15000 K. When we assume a temperature of $10^6$\,K, motivated by 
plausible shock velocities, the thermal pressure in these regions is 
still only $\approx$0.1-1 times the amount in the emission line regions 
(see Table\,\ref{tab:ne_temp_bubbles}). For the bubbles to be rapidly expanding
into the surrounding material, bulldozing it, they need to have higher
pressure than their surroundings. Over a large range of temperatures,
the pressure in the bubbles is simply not large enough for the amount of
expansion that would result in arc creation and shock ionization.

\section{Properties of the Nuclear Outflow}
\label{sect:whittle_quant}

Our spectrophotometric analysis of Mrk\,573 indicates that the emission
line regions are likely photoionized by the nuclear source. If there is
a shock present, it is not strong enough to form the material into arcs
and ionize it, agreeing with the conclusion of \citet{ferruit99}. We
now complement these analyses with a detailed investigation of the
emission-line region and outflow energetics, quantifying the outflow's
interaction with the host galaxy ISM. Specifically, we estimate the
energies and pressures of the emission-line regions (section
\ref{sec:pressure,etc}) and relate them to values calculated for the
outflow (section \ref{sec:jetprops}). These analyses broadly
follow the techniques applied by the \citet{whittle08} study of Mrk\,78,
and we shall adopt similar notation. By comparing the values of the
different energy components of the outflow and emission line regions, we can
confirm that the emission line regions are photoionized. The various
energies also indicate the importance of mechanical feedback, and the
overall strength of the outflow present (section \ref{sec:jetstrength}). We
also compare our results to those of the analogous emission-line regions
studied by \citet{whittle08}, and demonstrate that Mrk\,573's
emission-line regions are very similar to what is observed in Mrk\,78.

\subsection{Emission Regions: Masses, Energies and Pressures}
\label{sec:pressure,etc}

To understand the relationship between ionized and neutral gas in the
emission regions, we estimate the values of the various energy
components. This is critical for determining what contributes to the
kinematics, ionization, and structure, in addition to the relative
importance of mechanical feedback on the host. We are able to calculate
the energy of the various components by building off of determinations
of the flux, mass and lifetimes of the emission line regions. 

We assume that the portion of a given emission-line region within the
slit is representative of the larger region as a whole, and so multiply
the flux within the slit by the ratio of the total region area
($\theta_{ex}\times\theta_{ey}$) to the slit region area to approximate
the total flux.  The slit regions are the width of the slit, 0.2\arcsec\, 
multiplied by the size in arcseconds each region covers in the pure emission-line 
profiles (see sec.\,\ref{sec:extract}, Table\,\ref{tab:regional_feat_fluxes}). 
The physical sizes of the regions that we use are listed in
Table\,\ref{tab:jet_nums}.  Our region sizes and fluxes are smaller than
those measured by \citet{ferruit99} because we can better distinguish
the sharp boundaries with our improved resolution; \cite{ferruit99} had
an angular resolution of $\approx$0\farcs35 FWHM compared to our STIS
resolution of 0\farcs1 FWHM.

The ionized gas mass in each emission-line region ($M_{em}$) may be
estimated from the H$\beta$ emission-line luminosity and the electron
density $n_e$ derived from the [\ion{S}{2}] doublet ratio from
recombination theory:
\begin{equation}
M_{em} \approx \frac{L_{H\beta} m_p}{h\nu_{H\beta}n_e\alpha_{H\beta}}
\approx 8\times10^{14}\ \Big(\frac{cz}{H_o}\Big)^2 F_{H\beta} (n_{em,3})^{-1}
{\rm \ M}_\odot
\end{equation}
In this and subsequent equations, the numerical subscript appended to
a variable indicates a logarithmic scaling, e.g. $n_{em,3}$ represents
density in units of $10^3$ cm$^{-3}$.
The interaction age of the region, $t_{em}$, is estimated as the
crossing time, $t_{cross}$, the amount of time required for a compact
parcel of gas expanding outwards at $V_{em}$ to traverse the distance
from the nucleus to the arcs, i.e.
\begin{equation}
t_{em} \approx t_{cross} \approx 47100\ \Big(\frac{cz}{H_o}\Big)\theta_{ex} 
(V_{em,2})^{-1}\tan\phi\ \textrm{\ yr}
\end{equation}
where $\theta_{ex}$ is the angular separation between the nucleus and
arcs in units of arcseconds.
The emission-line masses for NW1 and SE2 are both $\sim 10000\pm
2200$\,M$_{\odot}$, while SE1 is less massive, $\sim 7000\pm
1500$\,M$_{\odot}$. SE1 has an interaction age of $3.53\pm1.3$\,Myr, 
and NW1 is $2.24\pm0.7$\,Myr old. SE2 has the largest interaction age, 
$t_{em}\approx6.90\pm2.5$\,Myr. This is the expected age progression, 
in that SE2 lies further from the nucleus than SE1 and NW1. However, 
we must note that SE2 does not appear to be strongly interacting with 
the biconical outflow, based on its lack of kinematic disturbance in 
H$\alpha$ (see Figure\,\ref{fig:halpha_rotcurve}). We use the interaction age 
of SE2 to determine other properties of the emission-line region 
to better understand the basic energetics of the region, 
but it is merely an approximation. The age determinations for all of 
the regions are consistent with what is seen in Mrk\,78 by \citet{whittle08} 
from emission-line structures of comparable scale (0.4--8.2\,Myr).

The total energy in the emission-line regions, $E_{em}$, may be
estimated from the product of the interaction age and the luminosity in
the [\ion{O}{3}]$\lambda$5007\AA\ emission line which measures the nebular
cooling power:
\begin{equation}
E_{em}= 1.2\times 10^{51}\ \Big(\frac{cz}{H_o}\Big)^2F_{5007}\times t_{em}
\end{equation}
where $F_{5007}$ is the [\ion{O}{3}] line flux in cgs units.
Each of the regions has total emission energy on the order of $10^{55}$\,ergs.
 
There are three energy components of interest in the emission line
regions associated with their mechanical and thermal energy content. The
translational kinetic energy, $E_{kin,t}$, of the outflowing gas may be
estimated from the mass of the region and its transverse velocity,
$V_{em}$,
\begin{equation}
E_{kin,t}\approx \frac{1}{2}M_{em}V_{em}^2\mathrm{cosec}^2\phi 
\approx 10^{53} M_{em,6}(V_{em,2})^2\mathrm{cosec}^2\phi\ 
\end{equation}
where we use the estimated bicone inclination angle $\phi$ to deproject
the observed radial velocity.  The internal (turbulent) kinetic energy,
$E_{kin,i}$, depends on the emission-line mass and velocity full-width,
$W_{em}$:
\begin{equation}
E_{kin,i}\approx \frac{1}{2}M_{em}(W_{em}/2.35)^2 
\approx 1.8\times10^{52} M_{em,6}(W_{em,2})^2\ 
\end{equation}
\citet{whittle08} used the width of the [\ion{O}{3}]$\lambda5007$\AA\ 
emission line, but because we have only low-dispersion spectra in which
these lines are either unresolved or marginally resolved at their
broadest, we instead use the widths of the H$\alpha$ and [\ion{N}{2}]
emission lines.  Finally the thermal energy, $E_{th}$, 
is
\begin{equation}
E_{th} \approx \frac{3}{2}NkT 
\approx 4.1\times10^{51} M_{em,6}T_{e,4}\ 
\end{equation}
where we use the electron temperatures derived from the [\ion{O}{3}]
emission lines.  All of these energies are listed in
Table\,\ref{tab:jet_nums}.  In general, the translational kinetic energy
in the regions is about 100 times larger than the thermal energy.  This
is consistent with our earlier suggestion that the gas is not
significantly shocked, or if shocked it has had enough time to cool
to the $\sim 10^4$\,K temperatures observed. This implies that the
acceleration mechanism must be fairly slow and gentle (the typical gas
speeds are $\sim 100-200$\kms\ in these regions, about an order of
magnitude larger than the sound speed for $10^4$\,K gas).  The internal
kinetic energy is roughly 10 times larger than the thermal energy, 
suggesting that this motion is not supersonic turbulence but
instead is bulk motion of the gas along the line of sight
that is unresolved at our scales, for example an expansion of order
$\pm100$\kms\ along the line of sight at the point of contact between
the nuclear outflow and the emission-line gas in the SE1 and NW1 arcs.

Our spectrophotometric analysis of Mrk\,573 suggests that the
emission-line regions are photoionized by the active nucleus rather than
ionized by fast shocks.  We investigate two sources of energy input into
these regions.  The first is the available photon energy, $E_{ph}$, from
the active nucleus.  We use the IRAS 60$\mu$m and 100$\mu$m flux from
Mrk\,573 to estimate the nuclear bolometric luminosity by assuming that
all photons from the central source are absorbed and re-emitted in the
infrared by surrounding dust, which underestimates the actual
luminosity.  Each region's $E_{ph}$ follows from the fraction of
nuclear energy that is intercepted by the arc, the covering fraction $cf$, 
over its interaction age ($t_{em}$) assuming that the regions have unity filling factor
\begin{equation}
E_{ph}= 1.5\times10^{39}\
\Big(\frac{cz}{H_o}\Big)^2\lbrack2.6S_{60}+S_{100}\rbrack\times t_{em} \times cf
\end{equation}
For the covering fraction, we assume that the depth of the cloud is approximately 
the arc thickness, $\theta_{ey}$. The depth could in fact be $\sim$10 times this size, 
increasing the photon energy by this same factor. For our purposes, being conservative
with our covering fraction is acceptable, as even with this small value, $E_{ph}$ 
is significantly larger than all other energy sources in the emission regions. 

The second source of energy is relativistic energy, $E_{rel}$, stored in
the radio outflow, which provides an estimate of the energy input due to
expansion of the radio lobes into the circumnuclear gas.  $E_{rel}$
depends on the minimum magnetic field strength, $B_{min}$, a lower limit
found by assuming equipartition between the relativistic particles and
the magnetic field
\begin{equation}
B_{min}\approx 2.93\times10^{-4}
\Big(\frac{S_{\nu}}{\theta_{rx}\theta_{ry}}\frac{a(1+z)^{3+\alpha_{r}}(\frac{30}{\lambda})^{\alpha_{r}}X_{0.5}(\alpha_r)}{f_{rel}\theta_{ry}(\frac{cz}{H_o})}\Big)^{ 2/7}
\end{equation}
\begin{displaymath}
\textrm{ where   }
X_q=\frac{(\nu_2^{q-\alpha_r}-\nu_1^{q-\alpha_r})}{q-\alpha_r}
\end{displaymath}
Where $\alpha_r$ is the
radio spectral index, and $f_{rel}$ is the filling factor. The spectral
indexes are observed to be $\alpha_{r}\approx -0.85$ for NW1
and $-0.5$ for SE1 \citep{falcke98}.  For the projected size of 
the radio outflow working surface we adopt $\theta_{rx}$ and $\theta_{ry}$ 
to be $\sim$0\farcs1. We adopt a$\approx$2, where a represents the contribution of 
relativistic energy density by ions in proportion to the contribution by electrons, 
following \citet{whittle08} and assume unity filling factor for the relativistic material ($f_{rel}=1$).  
Following our calculation of $B_{min}$, the
available relativistic energy, following \citet{whittle08}, is thus
\begin{equation}
E_{rel}\approx 1.6\times10^{56}\
\theta_{rx}\theta_{ry}^{2}f_{rel}^{3/7}\Big(\frac{cz}{H_o}\Big)^3B_{min}^2 
\end{equation}
The estimates for the NW1 and SE1 regions are listed in Table\,\ref{tab:jet_nums}.

Our calculations of the energies of the different components in the
emission line regions and the potential sources indicates that $E_{ph}$
dominates, being $\sim$1000 times larger than $E_{rel}$ and about 10 times
larger than the total energy in the emission regions, $E_{em}$. A
complete discussion of the different energy components follows in
sec.\,\ref{sec:jetstrength}.

The energetics of the emission line region and its possible 
radiative and relativistic sources are important for understanding 
the host-AGN interaction, and how it manifests itself. Similarly, 
examination of the different pressures can reveal whether the 
relativistic pressure from the radio lobe expansion or the 
radiative pressure from the central photon source are responsible 
for the arc structures. Following \citet{whittle08}, we 
calculate the emission-line region thermal pressure ($P_{em}$), the 
relativistic pressure ($P_{rel}$), and the radiation pressure ($P_{rad}$),
expressed in dynes cm$^{-2}$:
\begin{eqnarray}
P_{em}&\approx& 1.4\times10^{-9}\ n_{em,3}T_{em,4} \\
P_{rel}&\approx& 0.031\ B^2_{min}f_{rel}^{-4/7} \\
P_{rad}&\approx& 2.1\times10^{-11}\ \lbrack2.6S_{60}+S_{100}\rbrack\Delta_e^{-2}
\end{eqnarray}
where $\Delta_e$ is the distance from the region to the active nucleus. 
The thermal pressure, $P_{em}$, is based on the temperatures and
densities determined from our spectrophotometric analysis. The
relativistic pressure, $P_{rel}$, which results from radio lobe
expansion is based on the calculated magnetic field $B_{min}$. Lastly
the radiation pressure from the central photons, $P_{rad}$, is based 
on the FIR values, similar to our calculation of $E_{ph}$.  We find
$P_{rel}$:$P_{em}$$\approx$15:1, with $P_{rad}$ contributing
negligibly. In the next section we describe our methods for quantifying
the properties of the outflow itself, rather than its effect on the
emitting material. We combine these two analyses in
section\,\ref{sec:jetstrength}, and use the results to determine how
much influence the nuclear outflow of Mrk\,573 has on its host.

\subsection{Effect of an Outflow: Velocities and Dynamical Pressures}
\label{sec:jetprops}

Beyond analyzing the energies and pressures of the emission line
regions, we can quantify parameters of the outflow itself, in particular the
velocity and dynamical pressure. This allows us to determine the outflow's
strength directly, rather than conjecturing it based upon the behavior
of surrounding material. As the outflow does not appear to be affecting 
SE2, we only determine properties of it with respect to NW1 and SE1. 

The strength of the outflow is dependent upon its composition. A
relativistic outflow, in the form of a jet, with speed $V_o$ approximately 
equal to the speed of light, is able to both bulldoze and shock-ionize material. For an outflow 
with both thermal and relativistic material, the thermal component
literally weighs down the outflow, slowing it down to $\sim10^3$\,\kms\
\citep{whittle08}. Thus, outflow speed reflects the amount of thermal and
relativistic material present. \citet{whittle08} have suggested that the
outflow of Mrk\,78 is a heavy thermal jet, making it too weak to have
significant influence over its surroundings. Through our quantitative
analysis of the outflow in Mrk\,573, we can determine whether or not it
is also heavy and weak.

The approximate speed of the outflow is estimated from the energy and momentum 
values. An outflow with both material components can contribute to the 
kinetic and relativistic energy of the emission line regions; it can also 
affect the momentum of the emitting material, $G_{em}$. Thus, we use 
these values to determine $V_o$, the speed of the outflow. Following \citet{whittle08}:
\begin{equation}
\label{vel_eq}
V_{o}=2\frac{E_{kin}+E_{rel}}{G_{em}}\Big(1+\frac{1}{R_{kin}}\Big)^{-1} 
\end{equation}
where $R_{kin}$ is the ratio of kinetic energy to relativistic energy. 
If we assume the outflow primarily consists of heavy thermal material, 
making $R_{kin}\sim 1$, we calculate velocities of $\sim300-2000$\,\kms 
(see Table\,\ref{tab:jet_nums}), which is greater than the observed 
emission-line regions' speeds by factors of $\sim$10--100. This supports 
the suggestion that the outflow in Mrk\,573 is primarily made up of thermal 
material. A primarily relativistic outflow would have velocities vastly more 
than $\sim$10--100 times the emission-line velocities, and we do not 
calculate speeds of this extent. 

The thermal component of the outflow drives dynamical pressure, $P_{o,dyn}$,  into 
its surroundings. Following \citet{whittle08}: 
\begin{equation}
P_{o,dyn}=\frac{\Pi_o}{A_o}
\end{equation}
where $A_o$ is the area of the outflow, which we measure from the radio maps
of \citet{falcke98} to be $\approx 0.60$\,arcsecond$^{2}$, and we assume that $\Pi_o$, the force of 
the outflow, is $\approx \Pi_{em}$, the force in the emission line regions, 
calculated using the momentum and crossing times. We can then 
compare the ram pressure of the outflow to the pressures measured for 
the emission regions and evacuated bubbles, finding that it is significantly 
less than $P_{em}$ and $P_{rel}$. Following this, we use the dynamical pressure 
calculation to determine the Mach number of the outflow. 
\begin{equation}
N_{Ma}^2\approx \frac{3P_{o,dyn}}{5P_{rel}}
\end{equation}
As listed in Table\,\ref{tab:jet_nums}, the estimated Mach numbers,
though uncertain, are very small, on the order of 0.01, confirming that the outflow
is slow and thus, likely dominated by thermal material, similar to the outflow 
in Mrk\,78 \citep{whittle08}.

A final aspect of our outflow analysis is to determine the amount of mass 
the outflow transports. This value quantifies the influence of the 
outflow on its surroundings. The mass flux is:
\begin{equation}
\dot{M_{o}}=\frac{\Pi_o}{V_{o}}
\end{equation}
and is $\sim 10^{-11}$\,M$_{\odot}$\,s$^{-1}$ based on our estimates
above.

\subsection{Discussion: Outflow Strength}
\label{sec:jetstrength}
Our spectrophotometric analysis suggested that the outflow in Mrk\,573 is
unable to bulldoze material into the emitting arcs. Through analysis of
the physical properties of the emission regions and the outflow, we can
confirm this picture.

As shown in section\,\ref{sec:pressure,etc}, the relative proportions of
each of the energy components is \\
$E_{ph}$\,:\,$E_{em}$\,:\,$E_{rel}$\,:\,$E_{kin}$\,:\,$E_{th} \approx
10^4$\,:\,3000\,:\,10\,:\,1\,:\,0.01 (see Table\,\ref{tab:jet_nums}),
similar to what was found in Mrk\,78 by \citet{whittle08}.  The photon
energy E$_{ph}$ clearly dominates.  The emission-line region energy
E$_{em}$ is $\sim$25\% of E$_{ph}$, confirming that central
photoionization alone is sufficient to be the ionizing source of the
emission regions. The relativistic and kinetic energies, while both
substantially smaller than the emission-line region energy, are
comparable to each other. This implies that the slow expansion of the
radio source into the surrounding material is able to drive the small
amount of kinematic activity in the emission-line regions. Also note that 
SE2, where there is little evidence of interaction between the material and 
an outflow, has very similar energy magnitudes to SE1 and NW1. Despite the fact 
that NW1 and SE1 are in the direct path of the outflow, the regions appear 
quite similar to non-affected material, implying that an outflow has little 
influence on the emission-line regions' properties. 

We further investigate the kinematics by looking at the different
pressures in the regions. The radio outflow pressure $P_{rel}$ is
slightly larger than the ionized gas pressure, $P_{em}$, indicating that
the small kinematic motions in the emission regions are caused by the
gentle expansion due to the radio outflows.  Our previous analysis of
the densities and temperatures of the emission regions and the evacuated
bubbles shows that the bubbles have comparable or less thermal pressure
than the arcs over a wide range of temperatures. This supports the
conclusion that any expansion of the outflows is gentle. The radiative
pressure $P_{rad}$ is markedly less than the emission-line region
thermal pressure, and therefore makes no significant contribution to the
emission-line region kinematics we observe. We also investigate the
dynamic pressure contribution from the thermal component of the outflow, 
which has been suggested as a potentially significant pressure source \citep{bicknell98}. However for
Mrk\,573 the dynamic pressure is less than the relativistic
pressure. The dynamic pressure would have to be significantly larger than
the relativistic pressure or comparable to $P_{em}$ to be able to push
around the host ISM significantly. In this case, however, it appears
that the expansion is caused primarily by the radio lobes, with the mechanical 
contribution of the outflow being supplementary at best, and likely negligible.

Our calculation of the amount of material transported by the outflow also
indicates that it has little influence over the host. The outflow can
transport $10^{-4}$M$_{\odot}$\,yr$^{-1}$ of material. Over the
calculated lifetimes of the outflows, they have carried at most a 
tenth of the emission mass. Strong jets have been measured carrying as
much as $\sim$0.6\,M$_{\odot}$\,yr$^{-1}$ \citep{bicknell98},
significantly more than we calculate here, once again pointing towards
the general impotence of the outflow in Mrk\,573.

The energy analysis shows that radio outflow contributes primarily to
the kinetic energy of the ionized region but not to its ionization,
while the pressure analysis shows that the outflow does not strongly
expand into the surrounding material. While these estimates have
significant uncertainties associated with them, the differences between
the estimated energies, pressures, etc. are large enough that we can
clarify the relationship between the outflow and host despite this
uncertainty.  Taken together, the data suggest that the host galaxy of
Mrk\,573 is photoionized rather than shock-ionized, and that the radio
outflow gently shapes the regions but does not otherwise contribute to
its ionization or heating significantly. This is similar to what \citet{whittle08}
found for Mrk\,78.

The extensive evidence that the emission regions of Mrk\,573 are
pre-existing structures photoionized by a nuclear source also support
that the outflow itself is weak and likely thermal. An outflow made up of
primarily thermal material is slower and therefore less able to bulldoze
and strongly shock surrounding material. Earlier work by
\citet{bicknell98} on the outflow in NGC\,1068 has found similar results, 
suggesting that the radio jets in Seyferts differ from those in radio galaxies and quasars
primarily in that they are dominated by thermal gas instead of light
relativistic material. Our analysis of the outflow speeds confirm that
Mrk\,573 has a slow outflow, orders of magnitude slower than one dominated
by relativistic material. This suggests that the outflow of Mrk\,573 is
similar to the weak heavy jet that is incapable of shocking material and being 
the primary ionization mechanism in Mrk\,78 \citep{whittle08}.

Combining the different facets of this quantitative analysis presents a
consistent picture of a slow, heavy outflow capable of gently shaping the
surrounding host ISM into the arcs we see, but not fast enough to
substantially heat, displace, or ionize that material.

\section{Summary}

We have examined the influence of the AGN in Mrk\,573 on its host galaxy
using high angular resolution spectrophotometry from \hst\ STIS.  These
spectra provide both kinematic and spectrophotometric measurements
from which we derive the kinematics and physical state of the gas in the
circumnuclear regions associated with an interaction between a nuclear
radio outflow and spiral dust arms in the host galaxy.  Arguments from
emission-line diagnostics and thermodynamics (energy and pressure in the
various components) lead us to conclude that the outflow from the active
nucleus in Markarian\,573 does not strongly influence the surrounding
ISM other than gently sculpting the material into arcs.  The dominant
source of the heating and ionization in the gas can be entirely
explained by photoionization by energetic photons from the active
nucleus proper.  This is similar to what has been seen in other,
well-studied nearby AGN outflows with sufficient data to make similar
energetics arguments, particularly the work of Whittle et al. on Mrk\,78.
In many ways, Mrk\,573 and Mrk\,78 are very similar energetically and
morphologically: they have hot, heavy outflows sculpting extended
emission-line regions lit-up by ionizing photons from the nucleus.

All of the outflow and photoionization energetics taken together suggest
that while there is some feedback on the AGN's host, it is relatively
gentle and insufficient to unbind the host ISM or otherwise shutdown
star formation by many orders of magnitude. The outflows expand gently
into the host and have transported at most a tenth of the mass in the
emission regions over their lifetimes. The star formation regions in the
inner few kiloparsecs of Mrk\,573, seen as wound spiral arms outlined by
\ion{H}{2} regions \citep{pogge95}, appear to be undisturbed by the
outflow. This has important consequences for AGN feedback models in
galaxies. Specifically, this shows that outflows such as the one in
Mrk\,573 are not strong enough to remove material from the host galaxy;
in fact, analyses of low luminosity AGN indicate that AGN feedback has
little influence over the host
\citep{krongold07,whittle02,whittle05, whittle08}. As feedback 
is frequently invoked to explain different aspects of host-AGN evolution, new 
models must take into account that feedback, at least for AGN comparable to 
local Seyferts, has a significantly smaller effect on the host than previously thought.

\acknowledgements
We would like to thank Mark Whittle for useful discussions. 
Support for this work was provided by NASA through grant number GO-9143 from 
the Space Telescope Science Institute, which is operated by the Association of 
Universities for Research in Astronomy, Inc., under NASA contract NAS5-26555.

\begin{deluxetable}{lcccc}
\tabletypesize{\small}
\tablewidth{0pt}
\tablecaption{Line Flux Measurements \label{tab:regional_feat_fluxes}}
\tablehead{
\colhead{Feature} & \colhead{NW 1} & \colhead{NUC} & \colhead{SE 1} & \colhead{SE 2}}
\startdata
Aperture Area (arcsec$^2$) & 0.051 & 0.061 & 0.041 & 0.041 \\
\protect{[\ion{Ne}{5}]}$\lambda$3426\AA & 5.59e-17 & 3.40e-15 & 4.04e-17 & 5.16e-17 \\
\protect{[\ion{O}{2}]}$\lambda$3727\AA  & 3.11e-16 & 1.51e-15 & 7.09e-16 & 1.28e-16 \\
\protect{[\ion{Ne}{3}]}$\lambda$3868\AA & 1.17e-16 & 1.73e-15 & 2.42e-16 & 5.75e-17 \\
\protect{[\ion{Ne}{3}]}$\lambda$3968\AA & 3.76e-17 & 7.80e-16 & 1.37e-16 & 3e-17     \\
H$\gamma$ $\lambda$4340 \AA             & 5.07e-17  & 8.48e-16 & 1.03e-16  & 3.98e-17 \\
\protect{[\ion{O}{3}]}$\lambda$4363\AA  & 1.10e-17 & 4.22e-16 & 3.47e-17 & 5.90e-18 \\
\ion{He}{2}$\lambda$4686\AA             & 2.87e-17 & 8.96e-16 & 5.33e-17 & 1.63e-17 \\
H$\beta$ $\lambda$4861\AA               & 1.08e-16 & 1.85e-15 & 2.53e-16 & 7.24e-17 \\
\protect{[\ion{O}{3}]}$\lambda$4959\AA  & 4.70e-16 & 7.94e-15 & 9.58e-16 & 3.10e-16 \\
\protect{[\ion{O}{3}]}$\lambda$5007\AA  & 1.29e-15 & 2.37e-14 & 2.80e-15 & 8.98e-16 \\
\protect{[\ion{O}{1}]}$\lambda$6300\AA  & 5.32e-17 & 5.37e-16 & 1.02e-16  & 1.81e-17  \\
H$\alpha$ $\lambda$6563\AA              & 4.48e-16 & 7.14e-15 & 9.76e-16 & 2.58e-16 \\
\protect{[\ion{N}{2}]}$\lambda$6583\AA  & 4.32e-16 & 3.57e-15 & 9.62e-16 & 1.21e-16 \\
\protect{[\ion{S}{2}]}$\lambda$6716\AA  & 1.28e-16 & 6.29e-16 & 2.69e-16 & 4.81e-17 \\
\protect{[\ion{S}{2}]}$\lambda$6731\AA  & 1.31e-16 & 8.64e-16 & 3.24e-16 & 4.3e-17   \\
\enddata
\centering
\tablecomments{The line flux measurements for the four emission-line regions in Mrk\,573. The fluxes have the units erg s$^{-1}$ cm$^{-2}$ and uncertainties of 
20$\%$ or less.}
\end{deluxetable}

\begin{deluxetable}{ccccc}
\tabletypesize{\small}
\tablewidth{0pt}
\tablecaption{Density and Temperature Regional Analysis\label{tab:ne_temp_fluxes}}
\tablehead{
\colhead{Feature} & \colhead{NW 1} & \colhead{NUC} & \colhead{SE 1} & \colhead{SE 2}}
\startdata
$\frac{H\gamma}{H\beta}$       & 0.471 $\pm$0.13 & 0.458$\pm$0.13 & 0.407$\pm$0.11  & 0.549$\pm$0.15 \\
$\frac{\lambda\lambda4959+5007}{4363}_{obs}$ & 160.42$\pm$54 & 74.96$\pm$25 & 108.36$\pm$36 & 204.90$\pm$69 \\
$\frac{\lambda\lambda4959+5007}{4363}_{corr}$  & 160.59$\pm$54 & 77.17$\pm$26 & 125.53$\pm$42 & - \\
$\frac{\lambda6716}{\lambda6731}$            & 0.979$\pm$0.27  & 0.728$\pm$0.20 & 0.828$\pm$0.23 & 0.981$\pm$0.27 \\
$n_e$ (cm$^{-3}$)                 & 789$\pm$78   & 2437$\pm$243   & 1489$\pm$148   & 762$\pm$76 \\
$T_e$ (K)                       & 10814$\pm$1081  & 14454$\pm$1445 & 12445$\pm$1244  & 10046$\pm$1004 \\
\enddata
\tablecomments{The line ratios and derived densities and temperatures for the 
four extraction regions. The Balmer decrement for SE 2 was non-physical for the specified temperature, 
so we assume that the reddening is negligible. The uncertainties for the density and temperature are 10\%.}
\end{deluxetable}

\begin{deluxetable}{cccc}
\tabletypesize{\small}
\tablewidth{0pt}
\tablecaption{Emission Measure Analysis \label{tab:ne_temp_bubbles}}
\tablehead{
\colhead{} & \colhead{Temperature (K)} &\colhead{B$_{NW}$} & \colhead{B$_{SE}$} }
\startdata
I(H$\beta$) (erg s$^{-1}$ cm$^{-2}$ arcsec$^{-2}$) & -  & 4.1e-16  & 2.4e-16  \\
Depth (arcsec)                                  & - &  1.6     & 1.6      \\
Depth (parsec)                                  & - &  576     & 576      \\
Emission Measure (cm$^{-6}$ pc)                 & 15000 &  826     & 483      \\
$\langle n_{e} \rangle$ (cm$^{-3}$)             & 15000 &  1.2     & 0.9      \\
Emission Measure (cm$^{-6}$ pc)                 & $10^6$ & 36210    & 21197    \\
$\langle n_{e} \rangle$ (cm$^{-3}$)             & $10^6$ & 7.9     & 6.1     \\ 
\enddata
\tablecomments{Temperature and density calculations from the emission measure for the bubble regions as 
described in section \ref{sect:emission_measure}. B$_{NW}$ refers to the bubble between NW1 and the nucleus and B$_{SE}$ between 
the nucleus and SE 1. }
\end{deluxetable}

\begin{deluxetable}{lccccc}
\tabletypesize{\small}
\tablewidth{0pt}
\tablecaption{Quantitative Analysis of the Outflow Arcs\label{tab:jet_nums}}
\tablehead{
\colhead{Quantity} & \colhead{Symbol (units)} & \colhead{NW 1} & \colhead{NUC} & \colhead{SE 1} &
\colhead{SE 2}}
\startdata
Region Angular Size (x)   & $\theta_{ex}$    (arcsec) & 2.14$\pm$0.1 & 0.67$\pm$0.1 & 2.55$\pm$0.1 & 3.72$\pm$0.1  \\
Region Angular Size (y)   & $\theta_{ey}$    (arcsec) & 0.31$\pm$0.1 & 0.31$\pm$0.1 & 0.26$\pm$0.1 & 0.31$\pm$0.1  \\
Distance from Nucleus     & $\Delta_{e}$    (arcsec)  & 1.81$\pm$0.1  & - & 1.76$\pm$0.1 & 2.63$\pm$0.1 \\ 
Electron Temperature      & $T_e$            (K)     & 10814$\pm$1081 & 14454$\pm$1445 & 12445$\pm$1244 & 10046$\pm$1004  \\
Electron Density          & $n_e$            ($cm^{-3}$) & 789$\pm$78 & 2437$\pm$243 & 1489$\pm$148 & 762$\pm$76 \\
H$\beta$ Emission Flux    & $log(F_{H\beta})$ (erg s$^{-1}$ cm$^{-2}$) & -14.8$\pm$0.1 & -14.0$\pm$0.1 & -14.4$\pm$0.1 & -14.9$\pm$0.1  \\
\oiii $\lambda$5007 Emission Flux & $log(F_{5007})$   (erg s$^{-1}$ cm$^{-2}$) & -13.7$\pm$0.1 & -12.9$\pm$0.1 & -13.4$\pm$0.1 & -13.9$\pm$0.1 \\
Emission-line Width       & $W_{em,H\alpha}$    (km/s)  & 159$\pm$20 & 217$\pm$20 & 214$\pm$20 & 108$\pm$20   \\
Emission-line Velocity    & $V_{em,H\alpha}$    (km/s)  & 110$\pm$20 & - & -83$\pm$20 & -62$\pm$20  \\
Emission Mass             & $M_{em}$         ($10^{6} M_{\odot}$)$\pm$22$\%$ & 0.009 & 0.018 & 0.012 & 0.007  \\
Emission Region Age       & $t_{em}$         (Myr)   & 2.24$\pm$0.65 & - & 3.53$\pm$1.31 & 6.90$\pm$2.50 \\
Emission Luminosity       & $log(L_{em})$    (erg/s) & 41.1$\pm$0.1 & 41.9$\pm$0.1 & 41.4$\pm$0.1 & 40.9$\pm$0.1  \\
Emission Region Energy    & $log(E_{em})$    (erg)   & 54.97$\pm$0.15 & - & 55.46$\pm$0.18 & 55.26$\pm$0.18  \\
Emission Region Momentum  & $log(G_{em})$    (gm cm s$^{-1}$) & 44.81$\pm$0.15 & - & 44.81$\pm$0.18 & 44.45$\pm$0.18  \\
Emission Region Force     & $log(\Pi_{em})$  (dyne)  & 30.96$\pm$0.22 & - & 30.76$\pm$0.26 & 30.11$\pm$0.26 \\
Thermal Energy            & $log(E_{th})$    (erg)   & 49.60$\pm$0.10 & 50.03$\pm$0.10 & 49.79$\pm$0.10 & 49.46$\pm$0.10  \\
Transverse Kinetic Energy & $log(E_{kin,t})$ (erg)   & 52.06$\pm$0.15 & - & 51.94$\pm$0.18 & 51.45$\pm$0.18  \\
Internal Kinetic Energy   & $log(E_{kin,i})$ (erg)   & 50.61$\pm$0.10 & - & 51.00$\pm$0.10 & 50.17$\pm$0.10  \\
Total Mechanical Energy   & $log(E_{mec})$   (erg)   & 52.07$\pm$0.18 & - & 51.99$\pm$0.20 & 51.47$\pm$0.20   \\
Photon Energy             & $log(E_{ph})$    (erg)   & 55.6$\pm$0.1 & - & 55.9$\pm$0.2 & 56.1$\pm$0.2 \\
Relativistic Energy       & $log(E_{rel})$   (erg)   & $\approx$52$\pm$0.17 & - & $\approx$53$\pm$0.17 & - \\
Emission Region Pressure  & $log(P_{em})$    (dyne cm$^{-2}$) & -8.92$\pm$0.04 & -8.31$\pm$0.04 & -8.59$\pm$0.04 & -8.97$\pm$0.04   \\
Radiation Pressure        & $log(P_{rad})$   (dyne cm$^{-2}$) & -10.5$\pm$0.06 & - & -10.5$\pm$0.06 & -10.9$\pm$0.04 \\
Relativistic Pressure     & $log(P_{rel})$   (dyne cm$^{-2}$) & -8.0$\pm$0.14 & - & -7.6$\pm$0.14 & - \\
Outflow Ram Pressure      & $log(P_{o,dyn})$ (dyne cm$^{-2}$) & -10.9$\pm$.22 & - & -11.1$\pm$.26 & - \\
Outflow Speed             & $V_{o}$          (km/s)  & 340$\pm$228 & - & 1700$\pm$1207 & - \\
Outflow Mach Number       & $N_{Ma}$                 & 0.03$\pm$0.5 & - & 0.01$\pm$0.5 & -  \\
Magnetic Field Strength   & $log(B_{min})$   (Gauss) & -3.25$\pm$0.07 & - & -3.05$\pm$0.07 & - \\
IRAS 60$\mu$m Flux        & $S_{60}$         (Jy)    &  & 1.27$\pm$0.15 \\
IRAS 100$\mu$m Flux       & $S_{100}$        (Jy)    &  & 1.26$\pm$0.16   \\
\enddata
\tablecomments{A quantitative analysis of Mrk\,573 calculated based upon the 
formulations of \citet{whittle08}. These quantities are derived in section
\ref{sect:whittle_quant}. Uncertainties are calculated using standard error
propagation. }
\end{deluxetable}

\end{document}